\documentstyle[preprint,prb,aps]{revtex}

\begin{document}

\draft

\preprint{\today}

\title{Optical Properties of MFe$_4$P$_{12}$ filled skutterudites}

\author{S.V.~Dordevic$^{(1)}$, N.R.~Dilley$^{(1)}$, E.D.~Bauer$^{(1)}$,
D.N.~Basov$^{(1)}$, M.B.~Maple$^{(1)}$ and L.~Degiorgi$^{(2)}$} 

\address{ $^{(1)}$ Department of Physics, University of California, San 
Diego, La Jolla, CA 92093-0319}

\address{$^{(2)}$ Solid State Physics Laboratory HPF-F3/F18, 
ETH Z\"{u}rich, CH-8093 Z\"{u}rich, Switzerland}

\maketitle

\begin{abstract}

Infrared reflectance spectroscopy measurements were made on 
four members of the MFe$_4$P$_{12}$ family of filled skutterudites,
with M=La, Th, Ce and U. In progressing from M=La to U the system undergoes
a metal-insulator transition. It is shown that, although the
filling atom induces such dramatic changes in the transport properties of the
system, it has only a small effect on lattice dynamics. We discuss
this property of the compounds in the context of their possible
thermoelectric applications.

\end{abstract}

\narrowtext
  
\section{INTRODUCTION}

Filled skutterudites \cite{jeitschko77} are a family of compounds 
with the general structural formula MT$_4$X$_{12}$ (M=alkaline earth, 
rare earth, actinide; T=Fe, Ru, Os; X=P, As, Sb). Previous studies have
shown that their physical properties are in large part determined
by the M atoms. A variety of interesting physical phenomena, including
superconductivity, antiferromagnetism, fluctuating valence,
and a metal-insulator transition have been observed among the various
members of the filled skutterudites
\cite{meisner85,toricashvili87,delong85,meisner81,shirotani97,dilley98}.
Recently these compounds 
attracted renewed attention due to their potential application 
as thermoelectric materials \cite{mahan97,slack94,nolas96}.
Mahan and Sofo \cite{mahan96} theorized that optimal thermoelectric
properties might be realized for materials which possess a sharp feature
in the electronic density of states, which they modeled as a Dirac
delta function. Two different approaches have been used in attempts
to realize this situation experimentally.
One is by reducing the dimensionality of certain materials, as in
PbTe/Pb$_{1-x}$Eu$_x$Te multiple quantum wells \cite{dresselhaus98}.
The other approach, which we discuss here, explores
electronic f-orbitals of lanthanide atoms, that are almost completely
localized in many lanthanide-containing compounds and which
give a contribution to the density of states which is a Lorentzian of
very narrow width. In searching for materials that satisfy the requirements
outlined by Mahan and Sofo, we studied
four members of the MFe$_4$P$_{12}$ family with M=La, Th, Ce and U as
filling atoms.

\section{EXPERIMENT}

Single crystals of MFe$_{4}$P$_{12}$ compounds were grown in a molten tin
flux. High purity (99.9\% or better) pieces of the M elements
(La, Th, Ce or U), Fe, P and Sn were placed in a quartz tube in the
proportions of 1:4:20:50 and sealed under 150 mm of ultra-high purity argon
gas. The quartz tube was placed in a furnace and the temperature was elevated
to 1050-1150$^{o}$C. The quartz tube was maintained at 
this temperature for about a week, and the temperature was than decreased 
at the rate of 2 $^{o}$C/h. When the temperature of the furnace reached about 
600$^{o}$C, the quartz tube was dropped into water, and the product of the
reaction, consisting 
of several single crystals embedded in the Sn-P solution, was recovered.
These single crystals often grew in the habit of truncated octahedra typical
of skutterudite compounds and had dimensions somewhat less than
1x1x1 mm$^3$ (Fig.1, top panel). They could be isolated 
after etching the Sn-P matrix with a concentrated solution of HCl. 
The crystal structure of a few of these crystals was verified utilizing
Gandolfi or Laue back-reflection x-ray cameras.

Electrical resistivity $\rho_{dc}$(T) measurements were made using 
a standard four-probe technique. The shape of the crystals
prevented an accurate determination of the sample geometry and hence of the
absolute values of $\rho_{dc}$(T). Instead the scale of the electrical
resistivity data was determined from the room temperature value of the
resistivity inferred from the optical measurements.
For La and Th-based samples, $\rho_{dc}$ was extracted as the best fit to
the Hagen-Rubens formula $R(\omega)=1-\sqrt{2 \omega \rho_{dc}/\pi}$,
whereas for Ce- and U-based samples it was obtained
by extrapolating the optical conductivity down to zero frequency.

The electronic structure and charge dynamics in MFe$_4$P$_{12}$ were studied
using infrared and optical reflectance spectroscopy.
Near normal incidence reflectance $R(\omega)$ of 
single-crystal samples of MFe$_4$P$_{12}$ was measured at UCSD
in the frequency range 50-30,000 cm$^{-1}$ (approximately 5 meV - 3 eV).
Single crystal sample of LaFe$_4$P$_{12}$ is shown in the top panel of Fig.1.
In our experiments we have collected light reflected by one of the facets of
the crystal. To obtain the absolute value of R($\omega$)
samples were coated in-situ with gold or aluminum in the optical cryostat and
the spectrum of a metal-coated sample was used as a reference. This
procedure yields reliable absolute values of $R(\omega)$ and does not require
ambiguous corrections for diffuse reflectance \cite{homes93}.
Filled skutterudites provide an excellent illustration of 
the importance of performing optical measurements on
natural surfaces. The two La-based samples that were analyzed differed only
with respect to the condition of their surface: one of the surfaces was
polished, the other was not.
A scanning electron microscope (SEM) picture of the sample with
natural, unpolished surfaces is shown in Fig.1
(top panel) together with the far infrared (FIR) reflectance of both polished
and unpolished samples (bottom panel).
Obviously, the polishing procedure significantly damages the surface layer
of the sample, as evidenced by the following features in the reflectivity
data for the polished sample: (1) a lower overall free-electron reflectance
and (2) weaker peaks associated with phonons due to the lack
of crystalline long-range order in the surface layer.
Fortunately, the procedure for measuring reflectance described
above allows us to work with samples of submillimeter size, thus eliminating
the need to produce larger surfaces by polishing.
The measured frequency interval was extended up to 10eV at ETH on
polished specimens of LaFe$_4$P$_{12}$ and UFe$_4$P$_{12}$.

The complex conductivity
$\sigma(\omega)=\sigma_{1}(\omega)+i\sigma_{2}(\omega)$ 
and complex dielectric function 
$\epsilon(\omega)=\epsilon_{1}(\omega)+i\epsilon_{2}(\omega)$ 
were obtained 
from $R(\omega)$ using a Kramers-Kronig analysis. The
uncertainty of the $\sigma(\omega)$ and $\epsilon(\omega)$ spectra
due to both low- and high-frequency
extrapolations required for the Kramers-Kronig analysis are negligible in
the frequency range where the actual data exist.

\section{RESULTS}

Figure 2 shows the temperature dependence of the electrical resistivity
$\rho_{dc}(T)$.
The LaFe$_{4}$P$_{12}$ and ThFe$_{4}$P$_{12}$  compounds behave like metals,
having positive temperature coefficients of resistivity. The residual
resistivity ratios RRR $\equiv \rho(300 K) / \rho(4.2 K)$ for the two samples
are 94 and 1.8, respectively.
On the other hand, the resistivities of CeFe$_{4}$P$_{12}$ and
UFe$_{4}$P$_{12}$ have negative temperature coefficients
typical of insulators and semiconductors. The resistivities change
by as much as seven orders of magnitude between room temperature and 4.2 K.

Presented in Fig.3 are the raw reflectivity data at
room temperature for each of the compounds. As anticipated from transport
measurements, UFe$_4$P$_{12}$ shows insulating behavior.
The absence of the free electron background allows detailed analysis of the 
lattice dynamics. Another noticeable feature is relatively high reflectivity
($\sim$ 40 $\%$) in the mid-infrared region (500 - 5,000 cm$^{-1}$). In most
insulating compounds R($\omega$) does not exceed 10 $\%$ in this range.
We propose that this is a result of the high frequency contribution to the 
dielectric constant, $\epsilon_{\infty}$. This behavior seems to be common 
for all insulating skutterudites, and will be discussed further below.

For CeFe$_4$P$_{12}$, the other known nonmetallic member of the skutterudite
family, the reflectances of the two samples analyzed were the same at
frequencies $\omega>$300 cm$^{-1}$, but were quite different from each other
in the far infrared region.
Sample No.1 behaved like a semiconductor (some free electron background
is present), whereas sample No.2 was insulating. This sample-to-sample
variation was already evident in measurements of both transport and
thermodynamic quantities \cite{meisner85}. Since our optical experiments
probe the bulk of the crystal \cite{comment2} and results show significant
variations between samples, we conclude that this is a bulk effect.
Similar to UFe$_4$P$_{12}$, this material exhibits a high reflectance
($\sim$ 50 $\%$) in the mid-infrared part of the spectrum.
The increase in reflectivity in the near-infrared and visible is suggestive of 
interband transitions.

The third member of the family, ThFe$_4$P$_{12}$, displays
metallic behavior. The free-electron reflectance is much
higher here than in CeFe$_4$P$_{12}$ and UFe$_4$P$_{12}$,
but the relatively low plasma edge (around 3,500cm$^{-1}$) is
an indication of a low carrier concentration.
The increase of the reflectivity in the near-infrared and visible
parts of the spectrum is again indicative of interband transitions.

Finally, LaFe$_4$P$_{12}$ is also metallic and exhibits a 
far-infrared reflectivity that decreases as temperature increases
(data not presented in this paper). From transport measurements it is
known that this compound is a superconductor with a transition temperature
of around 4.2 K \cite{meisner81}. The general shape of the reflectivity
is the same as in ThFe$_4$P$_{12}$. The only difference between the two
is in the position of the "plasma minimum" in R($\omega$), which is shifted
to about 4,000 cm$^{-1}$ in LaFe$_4$P$_{12}$ (see Fig.3).

The next step in the analysis is to perform a Kramers-Kronig
transformation on the raw reflectivity
data. To do that one needs to extrapolate the reflectivity to the
regions where the actual data are not available.
For electrically conducting samples (LaFe$_4$P$_{12}$ and ThFe$_4$P$_{12}$),
high FIR reflectance suggests using a Hagen-Rubens (HR) extrapolation
(commonly used for metals):
$R(\omega)=1-\sqrt{2 \omega \rho_{dc}/\pi}$, 
where $\rho_{dc}$ is the only parameter of the model. The natural way to
proceed is to use $\rho_{dc}$ from transport measurements and calculate
$R(\omega)$. But since $\rho_{dc}$ was not available, we inverted the
procedure: we found the best fit to the reflectance data and used it
to calculate the absolute values of our transport measurements (Fig.2). 
The best fits for LaFe$_4$P$_{12}$ and ThFe$_4$P$_{12}$ were
85 and 200 $\mu \Omega cm$, respectively. A power law extrapolation
R($\omega$)$\sim$1/$\omega^{4}$ was used for high frequencies.
For the CeFe$_4$P$_{12}$ and UFe$_4$P$_{12}$ samples, we extrapolated the low
frequency part with a straight line: for the semiconducting CeFe$_4$P$_{12}$
sample perfect reflectance at zero frequency was assumed (suggested by
the tendency of the data), whereas for the insulating CeFe$_4$P$_{12}$ and
UFe$_4$P$_{12}$ compounds we assumed a constant reflectance (horizontal
straight line). For the high frequency part the same power law as before
was used.

Using these extrapolations, the Kramers-Kronig transformation was performed
and the frequency dependence for the complex conductivity $\sigma(\omega)$
and complex dielectric function $\epsilon(\omega)$
were obtained. Figure 4 shows results for the real part of the
conductivity $\sigma_1(\omega)$.
Characteristics anticipated from reflectance data are now
obvious: LaFe$_4$P$_{12}$ and ThFe$_4$P$_{12}$ are metallic,
one of the CeFe$_4$P$_{12}$ samples is semiconducting while the other is
insulating \cite{comment1}, and, finally, UFe$_4$P$_{12}$ is insulating.

Conductivity spectra for LaFe$_{4}$P$_{12}$ and ThFe$_{4}$P$_{12}$ are clearly
Drude-like, but we failed to obtain good fits to the formula
$\sigma_{1}(\omega)=\sigma_{0}/(1+\omega^{2}\tau^{2})$.
Instead, spectra were analyzed within the ``extended" Drude model.
Here one allows for the frequency dependence of the scattering
rate $1/\tau$. It can be shown \cite{tanner93} that causality
requires the effective mass to be frequency dependent as well.
The spectra of the quasiparticle effective mass $m^{*}(\omega)/m_{e}$
($m_{e}$ - electron band mass) and scattering rate $1/\tau(\omega)$
can be obtained from the optical conductivity:

\begin{equation}
\frac{m^{*}(\omega)}{m_e} = \frac{\omega_{p}^{2}}{4 \pi} \frac{\sigma_2(\omega)}
{\sigma_1^2(\omega)+\sigma_2^2(\omega)}\frac{1}{\omega}
\end{equation}

\begin{equation}
\frac{1}{\tau(\omega)}=\frac{\omega_{p}^{2}}{4 \pi} \frac{\sigma_1(\omega)}
{\sigma_1^2(\omega)+\sigma_2^2(\omega)}
\end{equation}
where the plasma frequency $\omega_p^2=4\pi n e^2/m_e$ 
is estimated from the integration of $\sigma_{1}(\omega)$
up to the frequency corresponding to the onset of the interband absorption.
When these equations are used to describe the response of a conventional
metal one finds that both the scattering rate and effective mass are
independent of frequency.
The results of the analysis for LaFe$_{4}$P$_{12}$
and ThFe$_{4}$P$_{12}$ are presented in Fig.5 (top and bottom panel,
respectively). For both compounds the effective mass (right panels)
is only slightly enhanced ({\it m$^{*}$/m$_{e}<$5}) and almost temperature
independent. The scattering rate (left panels) is linear in frequency
(up to 3000 cm$^{-1}$ for LaFe$_{4}$P$_{12}$ and
2000cm$^{-1}$ for ThFe$_{4}$P$_{12}$) and is temperature dependent.
The values obtained by extrapolating the curves down to zero frequency
($dc$ limit) are plotted in Fig.2 as full squares (for LaFe$_{4}$P$_{12}$)
and full circles (for ThFe$_{4}$P$_{12}$). Such a good scaling is not
surprising if one considers the relation $\rho_{dc}=m^*/ne^2\tau$ and
the fact that effective mass $m^*$ and carrier concentration $n$
are temperature independent. A huge factor of 50 difference in the residual 
resistivity
ratio RRR between La and Th-based samples is consistent with the 
temperature dependence of the scattering rate. At low
temperature the phonon contribution is strongly suppressed and the large
residual scattering in this region for ThFe$_{4}$P$_{12}$ indicates that 
the La-based sample was much ``cleaner".

Figure 6 displays the real part of the complex dielectric function
$\epsilon_1(\omega)$ up to 3,000 $cm^{-1}$ (0.3eV) for all the samples,
where contributions from the interband transitions at higher frequencies
are not shown. Also shown are the resonance plasma
frequencies $\omega_{p}^{*}$ (determined as zero crossings of
$\epsilon_1( \omega)$) for the
LaFe$_{4}$P$_{12}$, ThFe$_{4}$P$_{12}$ and semiconducting CeFe$_{4}$P$_{12}$,
and high-frequency contributions $\epsilon_{\infty}$ for the insulating
CeFe$_{4}$P$_{12}$ and UFe$_{4}$P$_{12}$.

\section{DISCUSSION}

\subsection{Electronic Properties}

Electronic properties of both filled and unfilled skutterudites have been
discussed in a number of publications
\cite{jeitschko77,grandjean84,danebrock96,evers95}. The filling atom is
always assumed to be positively charged and only weakly bound to the
transition-metal pnictogen polyanion. The bonds within the polyanion are
considered to be covalent.
A simple charge counting method is sometimes used to predict the
properties of the compounds. According to this method, the polyanion
[Fe$_4$P$_{12}$]$^{4-}$ should be isoelectronic with CoP$_3$
(Co$_4$P$_{12}$) which is semiconducting. The electronic properties of
MFe$_4$P$_{12}$ will therefore depend on the valence state of the filling
atom M. Metallic behavior is indeed observed in MFe$_4$P$_{12}$ where
M = La, Pr, Nd, Sm, or Eu is trivalent, and semiconducting behavior in
CeFe$_4$P$_{12}$ and UFe$_4$P$_{12}$,
where U and Ce may be tetravalent. However, there is growing consensus that
f-electron hybridization effects better explain the semiconducting behavior
in latter materials \cite{singh96,singh97,nordstrom96}, as we discuss later
in this paper. Furthermore, the charge counting arguments fail to predict
the correct behavior for
ThFe$_4$P$_{12}$, in which Th is believed to be tetravalent
\cite{danebrock96} and is therefore expected to make the compound
semiconducting (electronically saturated). Both the frequency (Fig.4)
and temperature (Fig.2) dependence of the conductivity
clearly demonstrate metallic behavior.
This simple charge counting method thus seems insufficient to
explain all the experimental results,
and more detailed band structure calculations are necessary.

The absolute values of the resistivity data for the metallic samples
(Fig.2, top panel)
are approximately two orders of magnitude higher than in typical metals.
Apparently the curve for the Th-based sample exceeds the value of
150 $\mu\Omega cm$, which is the so-called Ioffe-Regel limit.
In conventional metals it corresponds to the regime when the carrier
mean free path is approaching the lattice constant.
The value of 150 $\mu\Omega cm$ is calculated for conventional metals, where
$n\sim10^{22} $cm$^{-3}$, $\tau\sim10^{-14}$ s and $m^{*}/m_{e}=1$.
From the plasma frequency $\omega_p$ we estimated the carrier
concentration in the Th-based sample to be approximately two orders of
magnitude lower ($n\sim10^{20} $cm$^{-3}$) which places the Ioffe-Regel
limit for this compound much higher. The spectra of the dielectric constant
(Fig.6) of LaFe$_4$P$_{12}$ and ThFe$_4$P$_{12}$ also exhibit the
characteristics of
``poor" metals. The zero crossing of the dielectric constant, which
determines the plasma resonance frequency:

\begin{equation}
\omega_{p}^{*}=\frac{\omega_{p}}{\sqrt{\epsilon_{\infty}}}=
\sqrt{\frac{4\pi n e^2}{m_e \epsilon_{\infty}}}
\end{equation}
is in the range 0.1-0.3eV, confirming a low carrier concentration for both
compounds. 

Besides having low carrier concentration, these compounds also seem to
have an unusual frequency dependence of the carrier scattering rate:
$1/\tau\sim\omega$ (Fig.5, left panels).
We believe that this is caused by the interband transitions that are at
relatively low frequencies (around 10,000cm$^{-1}$, Fig.4)
and contribute considerably even below 1000cm$^{-1}$.
The interband transitions also affect the effective mass: they drive it below
the free electron mass, and even to negative values (Fig.5, right panels).

On the other hand, UFe$_4$P$_{12}$ and CeFe$_4$P$_{12}$  display all
the characteristics of insulators or semiconductors.
High values of $\epsilon_{\infty}$ (17 for UFe$_4$P$_{12}$,
31 for CeFe$_4$P$_{12}$ No.1 and
32 for CeFe$_4$P$_{12}$ No.2, see Fig.6) are typical for hybridization gap
semiconductors \cite{ogut96,damascelli97}, as explained later in the text.
This results in the anomalously large mid-infrared reflectance that was
mentioned above. Model calculations of reflectivity for several different
values of $\epsilon_{\infty}$ are presented in Fig.7. The other parameters of the
model, obtained as the best fit to the reflectance spectrum of the U-based 
sample, were kept constant. It is apparent that the increase of 
$\epsilon_{\infty}$ leads to the enhancement of the absolute value of
R($\omega$) in the mid-infrared.

All conductivity spectra show a broad maximum centered around 1.5eV (Fig.4),
a signature of an interband transition. At this moment, band structure 
calculations are available for Ce and La-based compounds only
\cite{singh96,singh97,nordstrom96}, and they do not show any characteristic
feature at this energy range. For CeFe$_{4}$P$_{12}$ a gap at the Fermi level of
about 0.34eV is predicted \cite{nordstrom96},
but our spectrum indicates a somewhat smaller value
of about 0.15eV. This value is also consistent with the activation energy
obtained from fitting resistivity measurements to an activation
conduction form, $\rho$=$\rho_{0}$exp($\Delta$E/k$_{B}$T), in the temperature
range 80K $<$ T $<$ 150K which yields an
activation energy $\Delta$E of around 0.13eV \cite{meisner85}.
One possible origin of the insulating behavior which has been previously
put forward \cite{meisner85} is hybridization between the f electron
orbitals present in UFe$_4$P$_{12}$ and CeFe$_4$P$_{12}$ with the
conduction electrons which opens a gap in the electronic density of
states at the Fermi energy. This would be analogous to the hybridization
gap seen in the intermediate valent systems SmB$_{6}$ and SmS
\cite{falicov81}, except that the latter materials have considerably
smaller activation energies than CeFe$_4$P$_{12}$.
Band structure calculations \cite{nordstrom96} also suggest that Ce $4f$ 
states hybridize strongly with Fe $3d$ and phosphorus $p$ states in
the vicinity of the Fermi level. The notion of hybridization is also
supported by the experimental findings presented
in this paper. Namely, hybridization gap semiconductors are predicted
\cite{ogut96} to have: (1) high values of $\epsilon_{\infty}$ and (2)
anomalously large values of Born effective charges (explained below).
We indeed find experimental evidence for both of these effects.
As La and Th do not have any f electrons, no hybridization takes place,
and thus LaFe$_4$P$_{12}$ and ThFe$_4$P$_{12}$ display metallic behavior.
Clearly, additional theoretical and computational work is required to
explain these experimental findings in metallic skutterudites.

\subsection{Phonon Spectra}

Filled skutterudites crystallize in the cubic structure 
(space group {\it Im$\overline{3}$ - T$_{h}^{5}$} \cite{lutz82}).
Applying the correlation method
\cite{fateley72}, the following decomposition into the irreducible
representations is obtained:
$\Gamma=2A_g+2E_g+4F_g+2A_u+2E_u+8F_u$.
Eight of these modes are infrared active (F$_u$ species), eight are Raman
active (A$_g$, E$_g$ and F$_g$ species) and the other are neither infrared
nor Raman active. All infrared active modes are triply degenerate.

Frequencies and oscillator strengths of all phonon modes identified in the
conductivity spectra are given in Table 1. The numerical values are
obtained as the best fits in the classical oscillator model
\cite{bruesch82}:

\begin{equation}
\sigma_{ph}(\omega)=
\frac{1}{4\pi}\sum_{j}\frac{S_{j}\gamma_{j}\omega^{2}}{(\omega_{j}^{2}-
\omega^{2})^{2}+\gamma_{j}^{2}\omega^{2}}
\end{equation}

Although the group theory analysis predicts eight infrared active modes,
we were able to identify only six from the spectra of the insulating samples.
The oscillator strength of the other two was probably below our detection
limits. Due to a high free-electron contribution to the conductivity,
only four phonons are identified in the Th-based sample and three
in the La-based sample. Interestingly, except for the lowest lying mode,
all peaks are at nearly the same frequency in all compounds. We propose
below that they are due to the Fe$_4$P$_{12}$ sublattice. The filling
ion M is believed not to be very strongly bound to this sublattice
and is free to rattle within the "cage". Weak binding could be the
reason for the high oscillator strength of this mode (Table.1)
The most interesting feature of the phonons is that
even though the conductivity (carrier
concentration) changes by almost three orders of magnitude going from the
U to La samples, the oscillator strength of the phonon modes is only weakly
affected (Table 1). Despite the fact that the conductivity of ``metallic"
samples is enhanced, the carrier density is still below 0.2 electrons/unit
cell which is much lower than the effective oscillating charge.

\subsection{Effective Charges}

The discussion in Section IV.A suggests that the metal-insulator transition
in the MFe$_4$P$_{12}$ series is in accord with the opening of a hybridization
gap in the electronic density of states. Another predicted feature of the 
hybridization gap semiconductors 
are anomalously high values of the effective dynamic charges \cite{ogut96}.
Our calculations of the Born dynamical charges $e_B^*(M)$ for
U and Ce-based samples are indeed in agreement with this prediction.
The Born effective charge can be determined from the formula:

\begin{equation}
\sum_{j}\left (\omega_{LO, j}^{2}-\omega_{TO, j}^{2}\right)=
\frac{Ze_{B}^{*2}(m)}{V_{B}c^{2}\pi\epsilon_{\infty}m}
\end{equation}
where $\omega_{LO, j}$ and $\omega_{TO, j}$ are frequencies of the
longitudinal and transverse optical phonons, respectively,
V$_{B}$ is the volume of the primitive cell, Z is the number of formula
units in the primitive cell (Z=2) and m is the reduced mass of the primitive
cell.

In accord with our assignment of phonons in the previous section,
we calculate $e_B^*$ for the filling ion and Fe$_4$P$_{12}$ sublattice
separately. For the former, the sum on the left-hand side of Eq.5 has only
one term (the lowest lying phonon), whereas for the latter it runs over
all the
remaining peaks. This procedure gives the following results
for the Fe$_4$P$_{12}$ sublattice: $e_B^*$=8.9 in UFe$_4$P$_{12}$ and 11 in
CeFe$_4$P$_{12}$. Such high values for a predominantly covalent lattice
(covalent bonding is expected within the Fe$_4$P$_{12}$ polyanion) indicate
that hybridization plays an important role.
The dynamical charge associated with the filling ion is also anomalously high.
For U and Ce-based samples we find $e_B^*$ of 8.9 and 8.7, respectively.
Obviously, such high values of the transverse dynamical charge
(i.e., the one that couples to the external radiation and that is measured in
optical experiments) cannot be due to the valence state of the ionically
bonded filling atoms.

\subsection{Thermoelectric Applications}

There has been an enormous amount of research in the field of 
thermoelectric materials in the recent years, with emphasis
on novel materials concepts. The figure of merit of a material
for thermoelectric applications is measured by the dimensionless quantity: 

\begin{equation}
ZT=\frac{\sigma S^{2}}{\kappa_{e}+\kappa_{l}} T
\end{equation}
where T is temperature, $\sigma$ is the electrical conductivity,
S is the Seeback
coefficient and $\kappa_{e}$ and $\kappa_{l}$ are electronic and
lattice contributions
to the thermal conductivity. For the novel TE materials to be
competitive with
conventional cooling systems, ZT $\approx$ 3 is required \cite{mahan97}.
The filled skutterudites (e.g., LaFe$_{3}$CoSb$_{12}$ and
CeFe$_{3}$CoSb$_{12}$) attracted attention as potential
thermoelectric materials because it was shown that they
exhibit large Seebeck coefficients typical of semiconductors yet have
moderately high electrical conductivity and also very low lattice thermal
conductivity $\kappa_l$ \cite{mahan97,nolas96,chen97}. The latter property
has been attributed to the localized, incoherent vibrations (rattling)
of the filling ions within the atomic "cages" formed by the other atoms.
The hypothesis was based on ultrasound, neutron diffraction, and specific
heat measurements,
and on the fact that $\kappa_l$ for the unfilled skutterudite CoSb$_3$
was several times larger at all temperatures \cite{keppens98}.

We believe that the
lowest lying phonon (which varies in frequency from compound to compound) 
is this rattling mode. All other phonons can be attributed to the dynamics 
of the Fe$_{4}$P$_{12}$ sublattice, which
explains their similarity in all four compounds (both in frequency and
in strength).
It has been recently suggested \cite{keppens98,mandrus97} that the rattling
of the filling ion can be treated as a localized vibrational mode.
We note that in UFe$_{4}$P$_{12}$ and CeFe$_{4}$P$_{12}$ the frequencies of
the lowest lying mode scale as $\sqrt{k/M}$, where M is the mass of the
filling ion and $k$ is the force constant, found to be the same in Ce and
U-based crystals. This scaling suggest that the filling ion is weakly bound
to the lattice. If the filling ion were strongly bound to the lattice, one
would expect the force constant to be quite different for
different ions. Based on the calculated value of the force constant
we estimate
the phonon frequencies for the same mode in ThFe$_{4}$P$_{12}$ and
LaFe$_{4}$P$_{12}$. They are expected to
be around 82 and 141cm$^{-1}$, respectively (indicated by arrows in Fig.3).
The free electron background in raw reflectance data 
in this frequency range for these two compounds
is too high to allow detection of any phonons.

In order to produce viable thermoelectric materials, careful tuning of
the compound's carrier concentration is required.
Several alloying schemes are possible within the filled
skutterudites. Compounds of the form
M(T$_{x}$T'$_{4-x}$)X$_{12}$, MT$_{4}($X$_{y}$X'$_{12-y}$) or even 
M(T$_{x}$T'$_{4-x}$)(X$_{y}$X'$_{12-y}$) have been successfully synthesized
\cite{mahan97,nolas96,chen97}.
Recently compounds of the form La$_{1-x}$Ce$_{x}$Ru$_4$P$_{12}$ 
($0\leq x \leq 1$) were
produced at high pressures \cite{shirotani99}, demonstrating the 
ability to vary $x$ continuously over the entire alloying range.
We believe that this may allow us to
design a material whose electrical and thermal conductivities are
optimal for thermoelectric applications.

Finally, we want to discuss our experimental results from the point of
view of Mahan and Sofo's theory \cite{mahan96}. They point out 
that all transport coefficients relevant for thermoelectric materials
($\sigma$, S and $\kappa_{e}$) are functions of 
the electronic density of states N$_{el}$(E). 
They find, in the case of a material with the minimum lattice
thermal conductivity $\kappa_l \approx 2 mW/(cm K)$, that a sharp feature
(theoretically a Dirac delta function) in the electronic density of states
located 2.4k$_{B}$T away from the Fermi level 
is needed to maximize ZT. This is equivalent to having a resonance
approximately 60meV (at room temperature) above or below the Fermi level. 
None of the materials we
have analyzed in this paper shows any characteristic feature on this energy
scale (Fig.4). We are currently studying another member of the
skutterudites family, YbFe$_{4}$Sb$_{12}$ \cite{dilley98},
and find interesting charge dynamics in the anticipated energy range 
\cite{dordevic99}.

\section{CONCLUSIONS}

We have presented evidence in this paper that transport and optical
properties of the filled skutterudites are primarily determined by the
filling ion. The experimental results support the idea that hybridization
plays an essential role in determining the properties of the
MFe$_{4}$P$_{12}$ compounds.
From the optical spectrum of samples with M = Ce and U, we inferred the
following properties:
(1) a gap in the electronic density of states, (2) a large value of
the dynamical effective charges, and (3) an enhanced dielectric constant
at high frequencies $\epsilon_{\infty}$.
Localized, incoherent vibrations (``rattling") of the filling ions
were successfully modeled as Einstein oscillators, in agreement with
recent reports \cite{keppens98,mandrus97,feldman99}. Atomic substitutions at
this site will not only tailor
the electrical conductivity of the material but may also increase
the phonon scattering due to the rattling of the various filling ions.
We believe that this method (atomic substitution at the filling atom site),
combined with alloying at the Fe
and P sites, opens possibilities for significant improvements of the
thermoelectric coefficient. Further research,
especially measurements of thermal conductivity,
is needed to verify these ideas.

\section{Acknowledgements}

Research at UCSD was supported by the U.S. Department of Energy under
Grant No. DE-FG03-86ER-45230, the U.S. National Science Foundation under
Grants No. DMR-97-05454 and DMR-98-75-980, the Campus Laboratory
Collaboration of the University of California and Sloan Foundation.
D.N.B is a Cottrell Fellow of the Research Corporation.

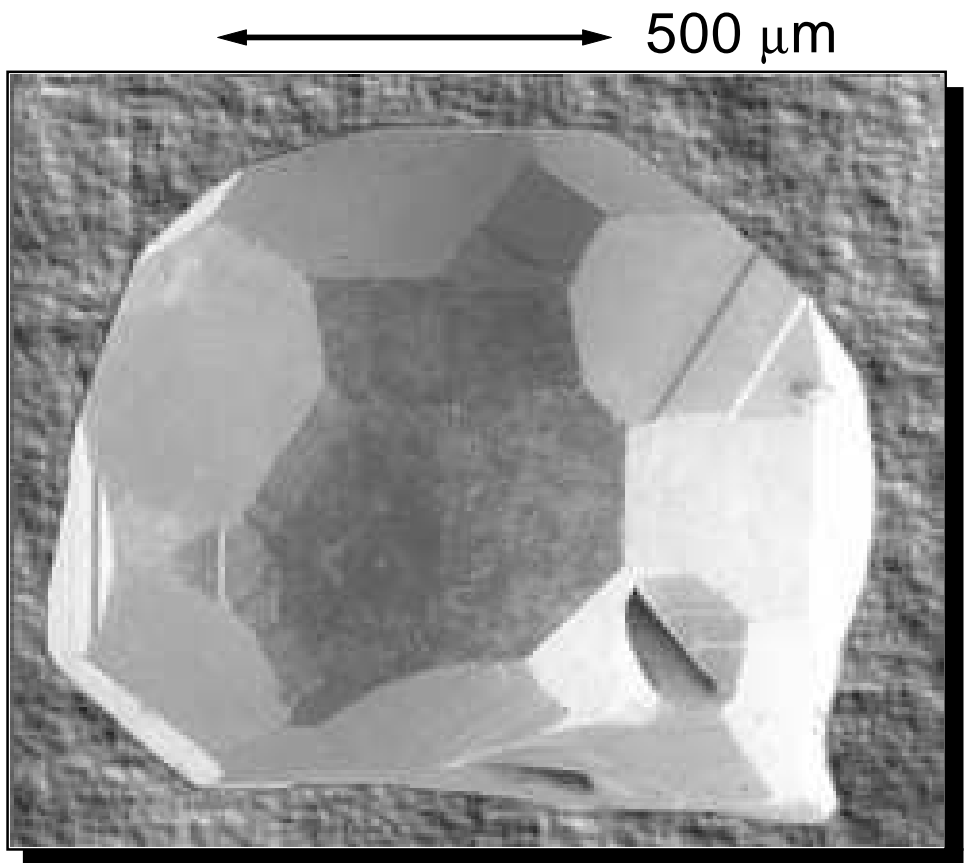
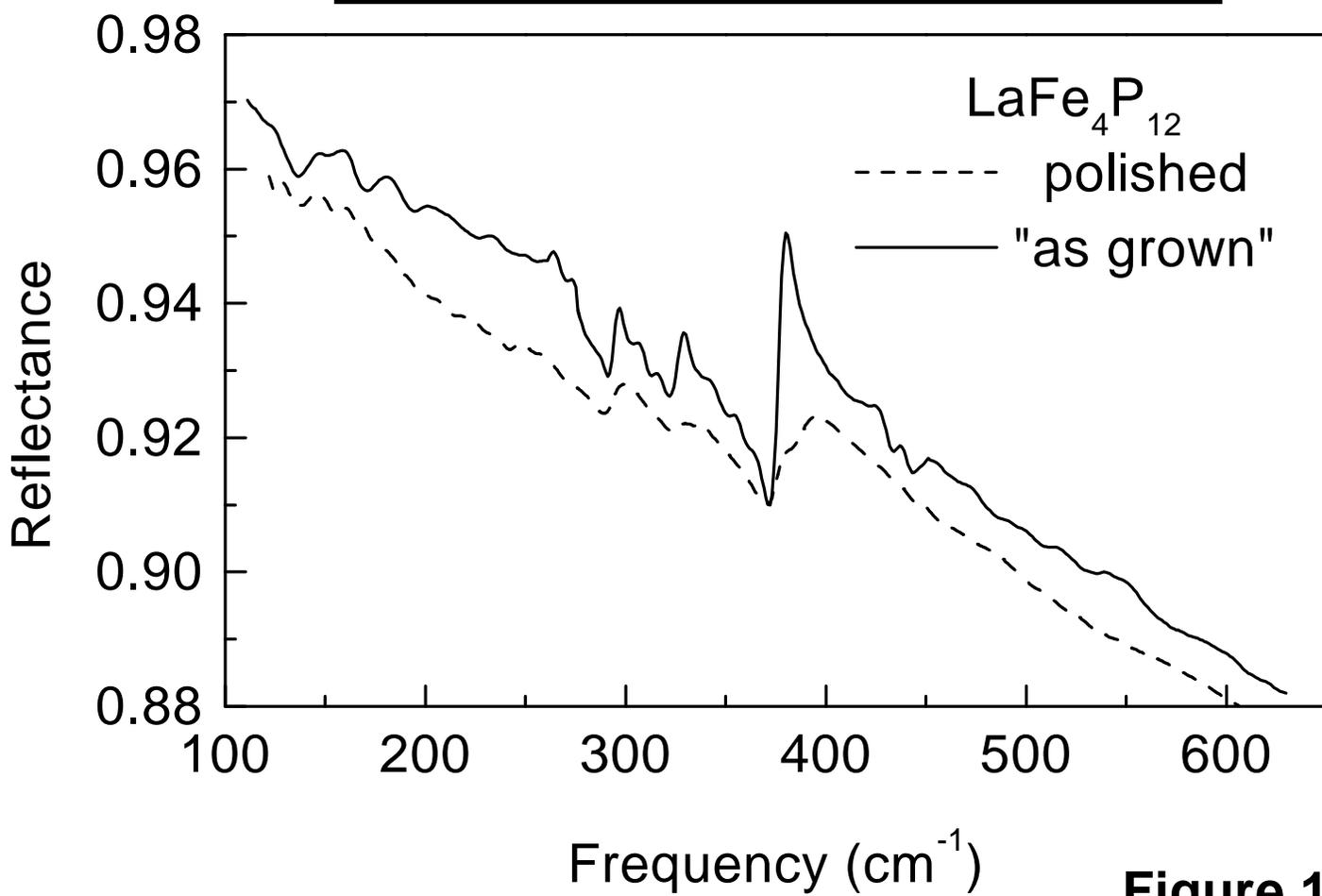
\begin{figure}
\caption{An SEM picture of as-grown LaFe$_4$P$_{12}$ single crystal 
(top panel).
Reflectance spectra (bottom panel) show the importance of natural surfaces;
polishing damages the surface layer of the sample, creating defects
that increase resistivity (lower the reflectance). The phonons are also much
weaker due to the lack of long range order.}
\end{figure}

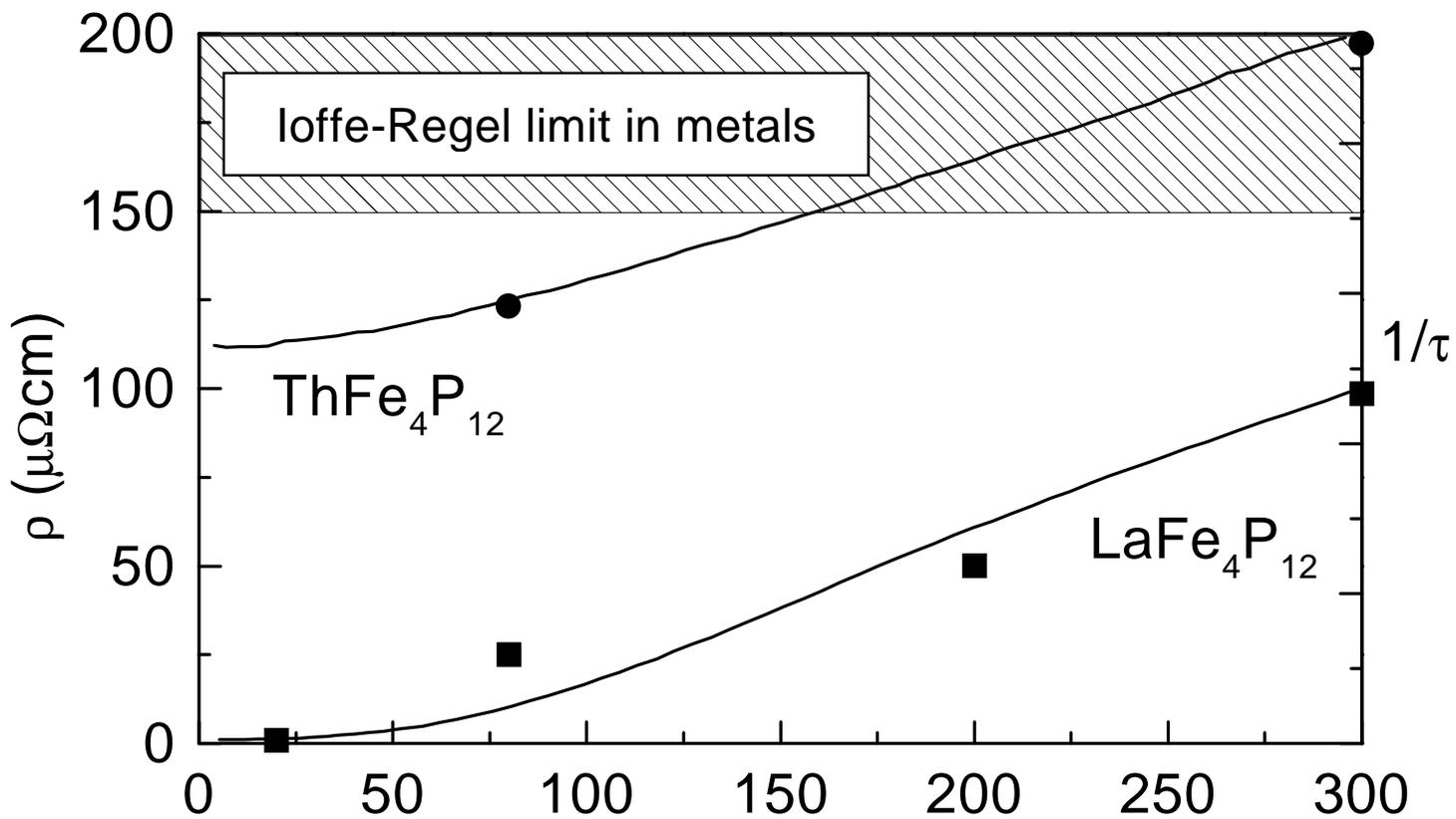
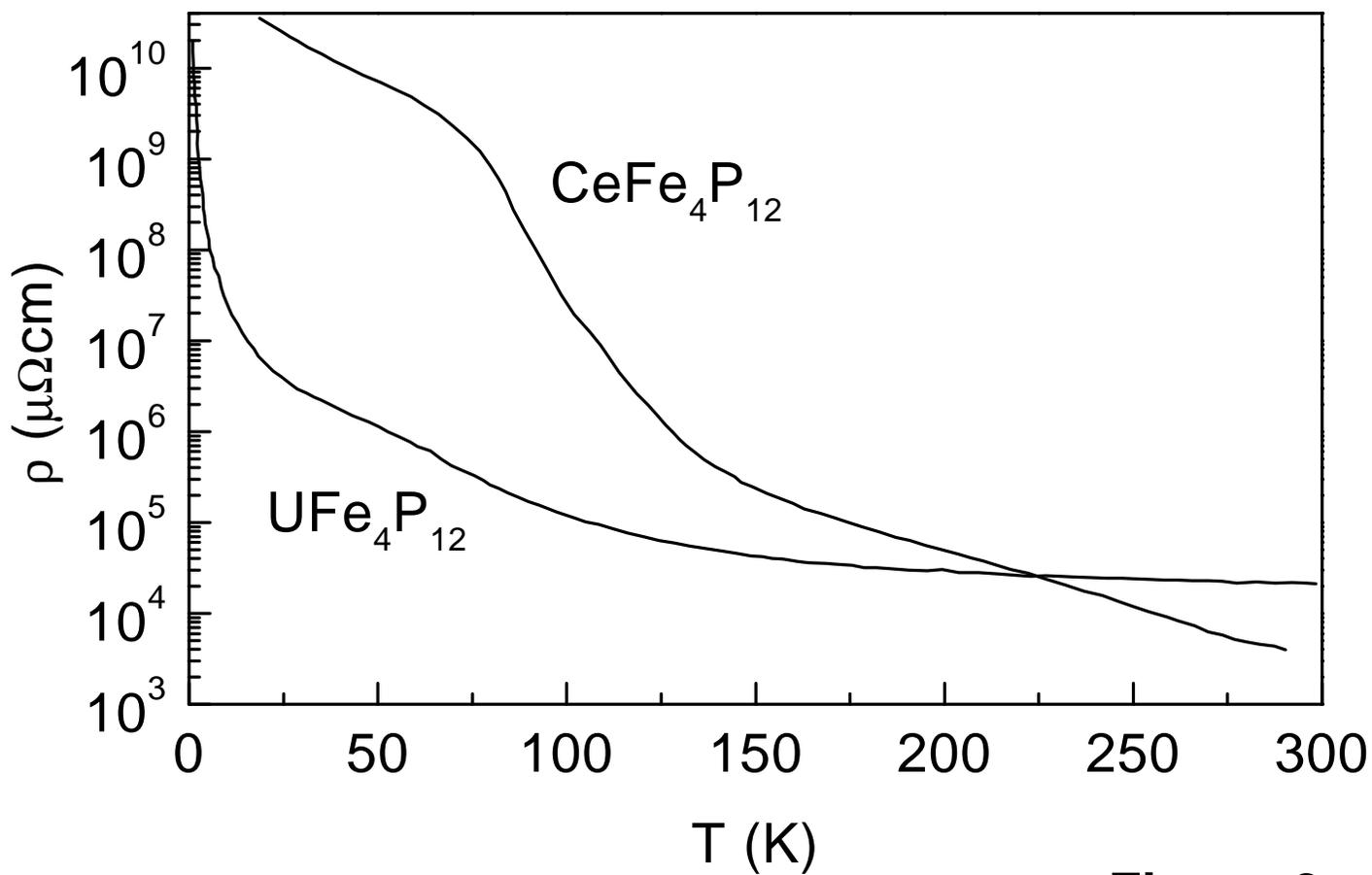
\begin{figure}
\caption{Temperature dependence of the $dc$ resistivities of MFe$_4$P$_{12}$.
The absolute values are obtained by extrapolating the optical
conductivity down to zero frequency (for U- and Ce-based samples) and from the
Hagen-Rubens fits to the reflectance (for Th- and La-based samples)
Also shown is the Ioffe-Regel limit calculated for
n$\sim$10$^{22}$ cm$^{-3}$ (typical for conventional metals).
In ThFe$_4$P$_{12}$ the carrier concentration
is two orders of magnitude lower, placing the Ioffe-Regel limit much higher.
The carrier scattering rate (right-hand axis in the top panel), is
shown as full circles for LaFe$_4$P$_{12}$ and full squares for
ThFe$_4$P$_{12}$. This values are obtained by extrapolating the 
scattering rate to zero frequency (see Fig.5).}
\end{figure}

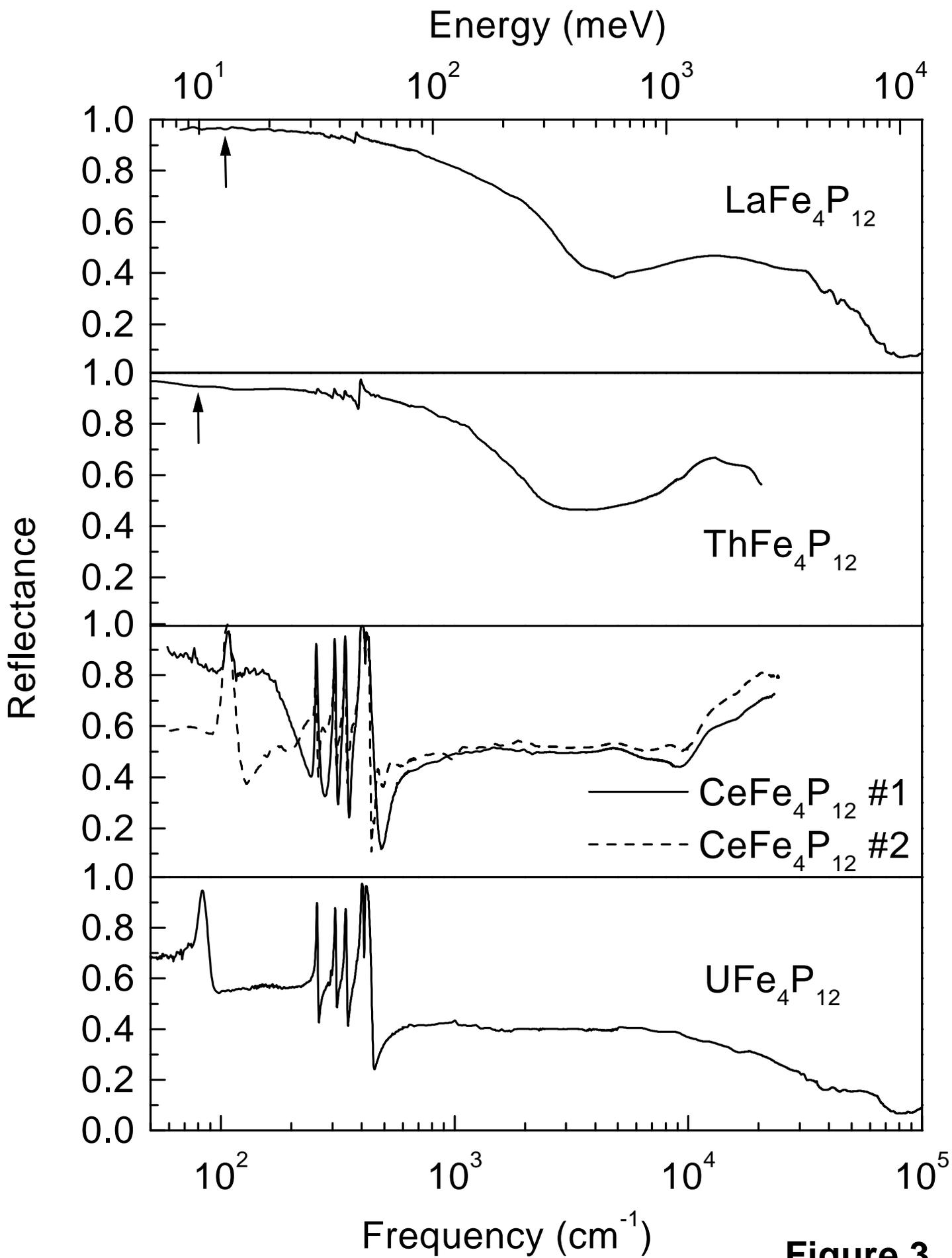
\begin{figure}
\caption{Reflectance spectra of four members of the skutterudite family.
The arrows on the graphs for La and Th samples indicate the positions
of the "rattling" mode, calculated from the simple harmonic model
(see text for details).}
\end{figure}

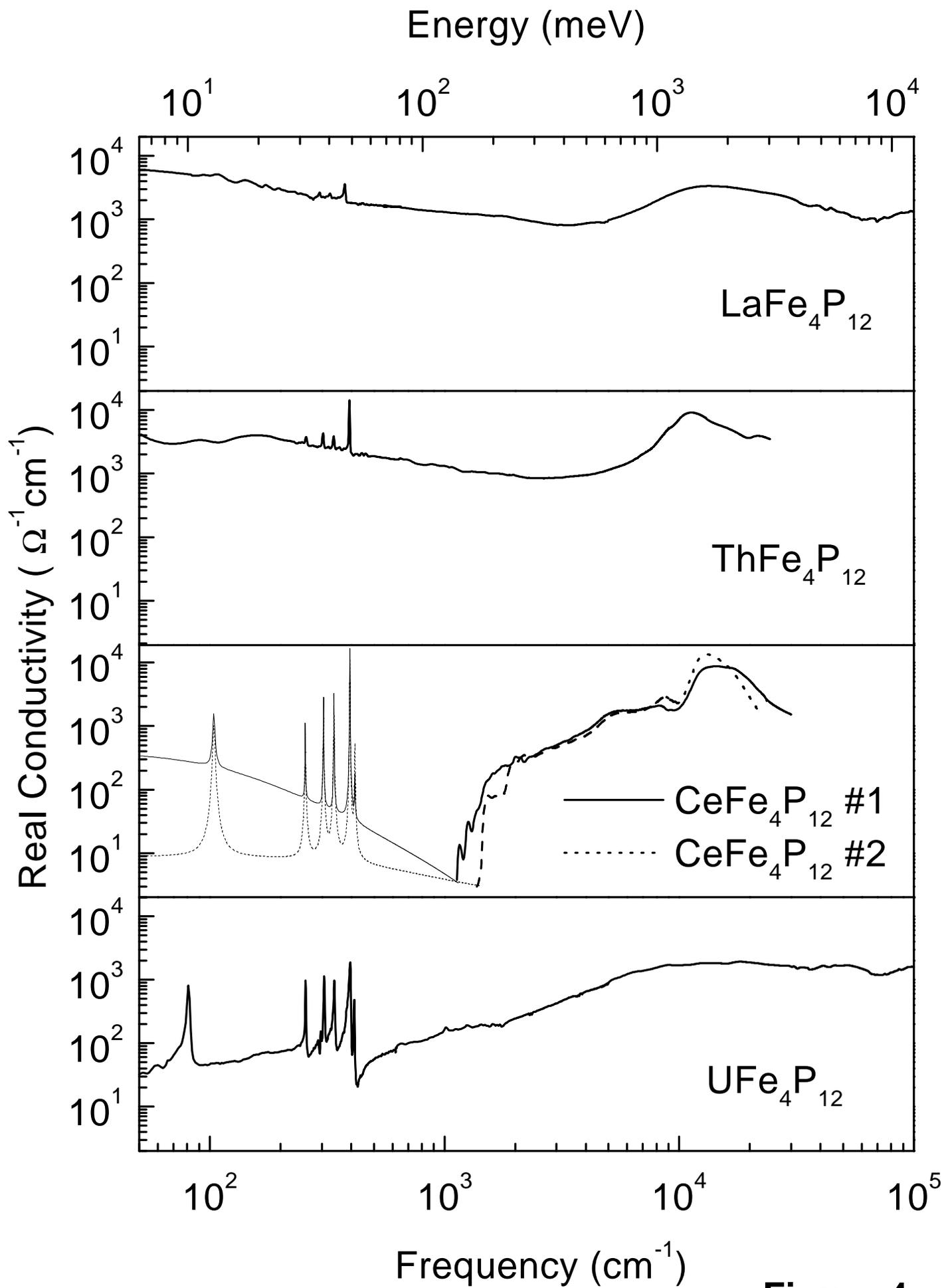
\begin{figure}
\caption{Real part of the optical conductivity as obtained from
a Kramers-Kronig analysis. All spectra show a broad maximum around 1.5eV,
which we attribute to an interband transition.
The low frequency part of the CeFe$_4$P$_{12}$ spectra (thin lines)
is obtained by fitting the reflectance data.}
\end{figure}

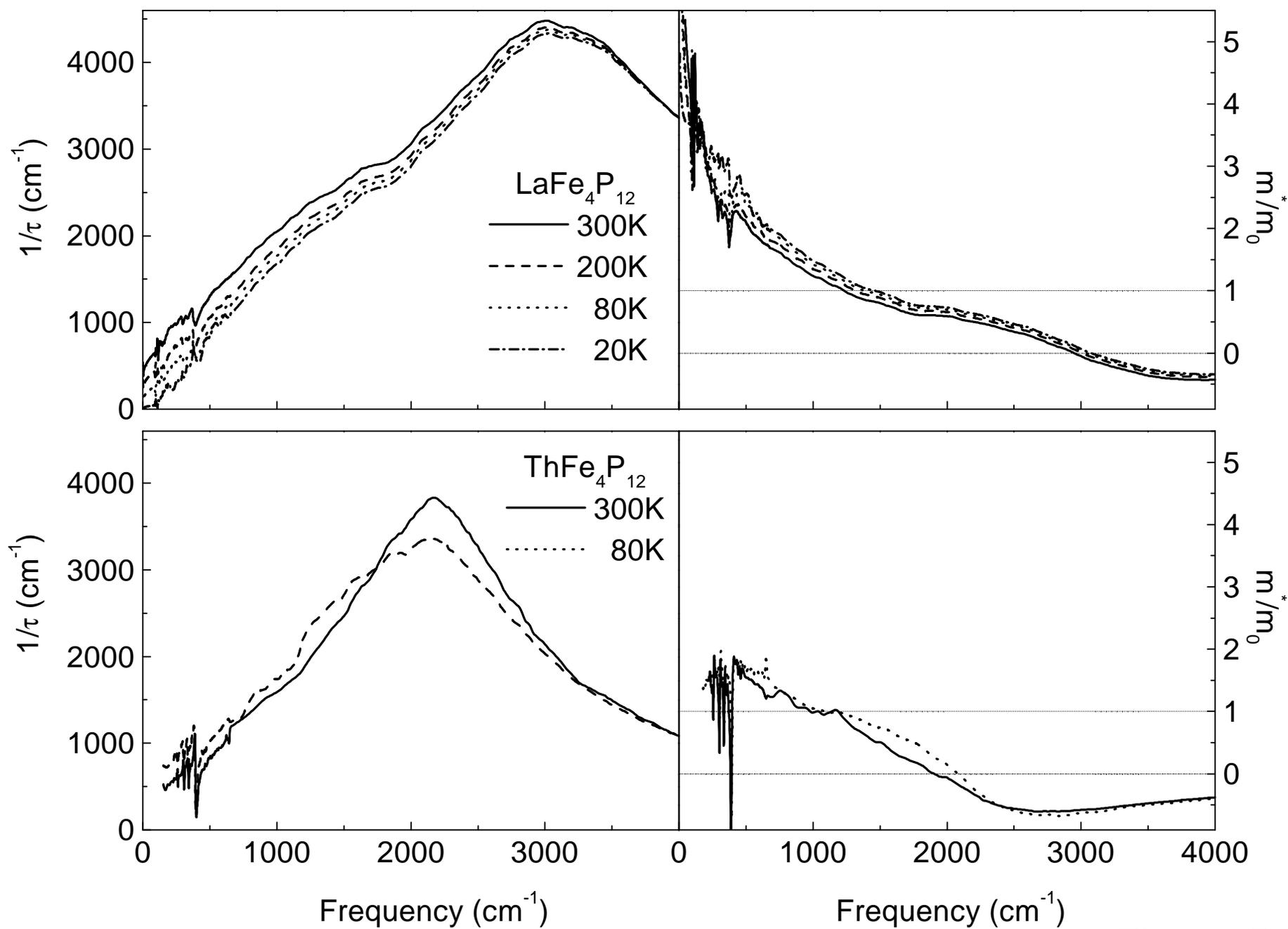
\begin{figure}
\caption{Frequency dependent effective mass (right panels) and scattering
rate (left panels) obtained within the ``extended" Drude formalism
of Eq.1 and Eq.2 (see text), applied to LaFe$_4$P$_{12}$ (top panels) and
ThFe$_4$P$_{12}$ (bottom panels). Note a slight mass enhancement and
linear frequency dependence of the scattering rate.}
\end{figure}

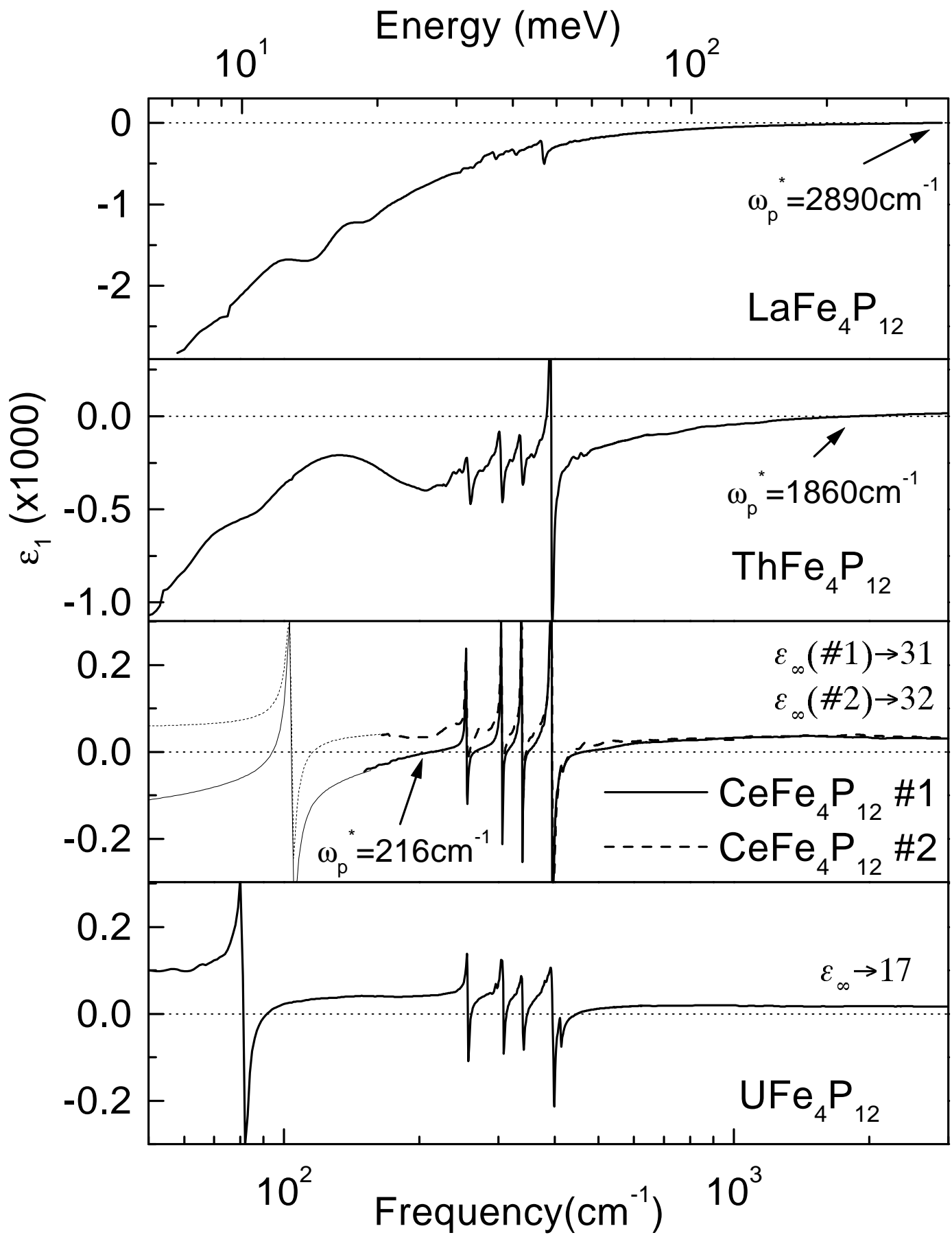
\begin{figure}
\caption{Real part of the dielectric function, shown over a narrower
frequency range (up to about 0.3eV). Again, thin lines are used for the
simulated region (for CeFe$_4$P$_{12}$). The low values of the plasma
resonance frequency $\omega_{p}^{*}$ are an indication of low carrier
concentrations. The high frequency contributions to the dielectric function
($\epsilon_{\infty}$) for insulating Ce and U-based samples are also shown.
High values are typical of hybridization gap semiconductors.}
\end{figure}

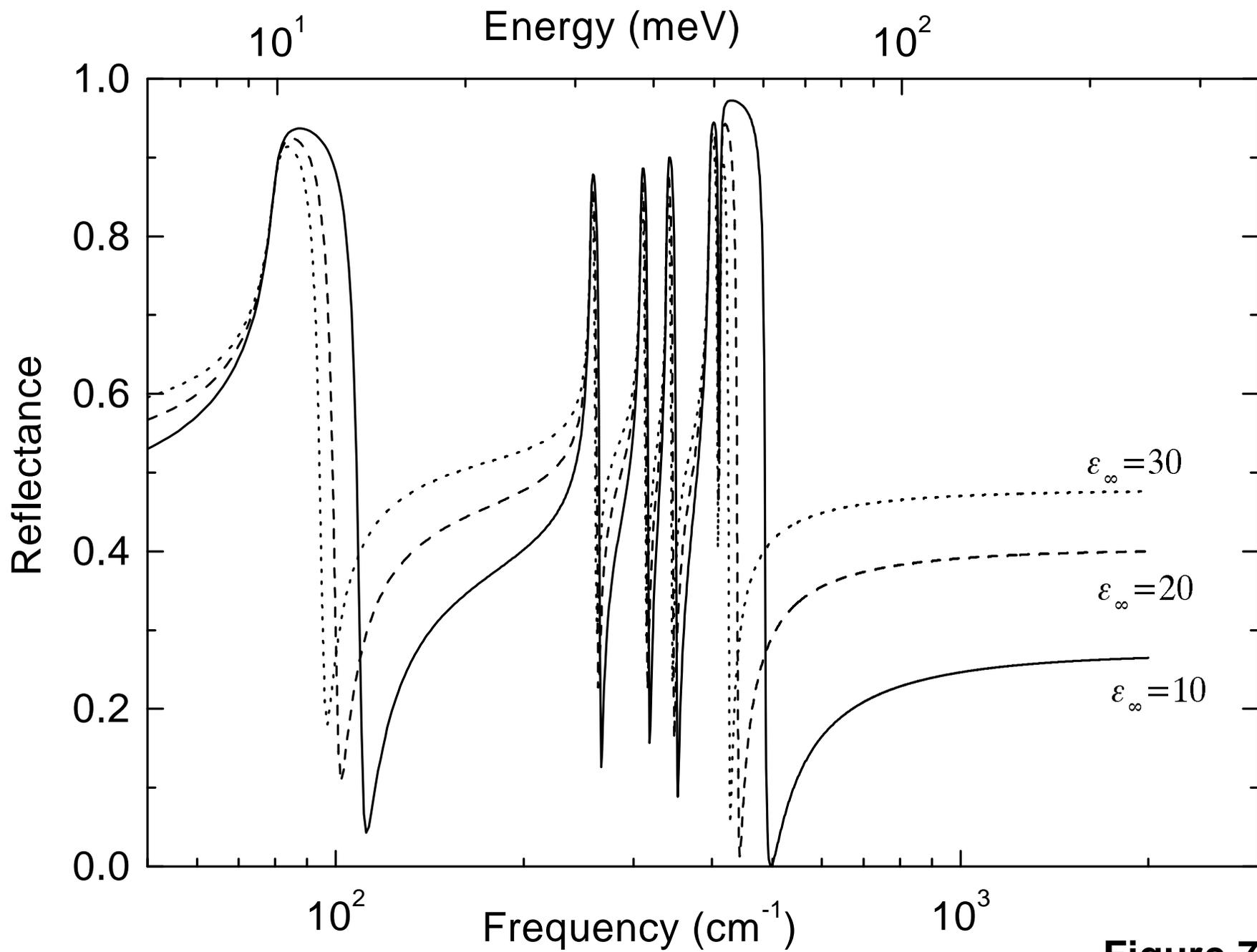
\begin{figure}
\caption{Model calculations of the reflectance for UFe$_4$P$_{12}$, shown
for several values of the free parameter $\epsilon_{\infty}$. 
The phonon parameters $\nu_j$ and $\rho_j$ for these curves are given in 
Table 1.
Apparently, an increase in $\epsilon_{\infty}$ leads to an increase in the 
mid-infrared reflectance.}
\end{figure}

\begin{table}
\caption{Frequency $\omega_j$ in cm$^{-1}$ and reduced oscillator strength
$\rho_{j}=S_{j}/4\pi\omega_{j}$ (in parentheses) of infrared
active phonon modes at {\bf q}$\approx${\bf 0} obtained by
a fit to the classical oscillator model.}
\centering
\begin{tabular}{|r|cccccc|}     \hline
                        & $\nu_{1}$& $\nu_{2}$& $\nu_{3}$& $\nu_{4}$& $\nu_{5}$& $\nu_{6}$ \\   \hline\hline
LaFe$_{4}$P$_{12}$      &          &         &294 (0.08)&325 (0.08)& 376 (0.22)&           \\   
ThFe$_{4}$P$_{12}$      &          &257 (0.14)&304 (0.20)&337 (0.11)& 393 (0.73)&           \\ 
CeFe$_{4}$P$_{12}$      &104 (1.17) &255 (0.07)&305 (0.14)&338 (0.13)& 396 (0.51)& 414 (0.01) \\
UFe$_{4}$P$_{12}$       & 80 (1.27) &256 (0.13)&308 (0.12)&339 (0.10)& 397 (0.22)& 411 (0.03) \\  \hline
\end{tabular}
\end{table}


\begin{references}

\bibitem{jeitschko77} W.~Jeitschko and D.~Braun, Acta Crystallographica,
Section B (Structural Crystallography and Crystal Chemistry)
{\bf B33} pt.11, 3401-6 (1977)

\bibitem{meisner85} G.P.~Meisner, M.S.~Torikachvili, K.N.~Yang, M.B.~Maple
and R.P.~Guertin, J.Appl.Phys. {\bf 57}, (1) 3073, (1985)

\bibitem{toricashvili87} M.S.~Torikachvili, J.W.~Chen, Y.~Dalichaouch, 
P.P.~Guertin, M.W.~McElfresh, C.~Rossel, M.B.~Maple and G.P.~Meisner,
Phys.Rev B, {\bf 36}, 8860-4, (1987)

\bibitem{delong85} L.E.~DeLong and G.P.~Meisner, Solid State Communications,
Vol.{\bf 53}, No.2, pp. 119-123, (1985)

\bibitem{meisner81} G.P.~Meisner, Physica {\bf 108B}, 763-4, (1981) 

\bibitem{shirotani97} I.~Shirotani, T.~Uchiumi, K.~Ohno, C.~Sekine,
Y.~Nakazawa, K.~Kanada, S.~Todo and T.~Yagi, Phys.Rev.B {\bf 56},
No.13, 7866-9, (1997)

\bibitem{dilley98} N.R.~Dilley, E.J.Feeman, E.D.Bauer and M.B.Maple,
Phys.Rev.B {\bf 58}, No.10, 6287-90, (1998)

\bibitem{mahan97} G.~Mahan, B.~Sales and J.~Sharp, Physics Today, 42-7,
March 1997

\bibitem{slack94} G.A.~Slack and V.G.~Tsoukala, J.Appl.Phys.,
{\bf 76} (3) 1665-71, (1994)

\bibitem{nolas96} G.S.~Nolas, G.A.~Slack, D.T.~Morelli, T.M.~Tritt and
A.C.~Ehrlich, J.Appl.Phys. {\bf 79} (8) 4002-8, (1996)

\bibitem{mahan96} G.D.~Mahan and J.O.~Sofo, Proc.Nat.Acad.Sci.USA, 
Vol.{\bf 93}, 7436-9, (1996)

\bibitem{dresselhaus98} T.Koga, S.B.Cronin, T.C.Harman,
X.Sun and M.S.Dresselhaus, Mater.Res.Soc.Symp.
Proc., Vol.{\bf 490}, 263-8, (1998)

\bibitem{homes93} C.~Homes, M.A.~Reedyk, D.A.~Crandels and T.~Timusk, 
Applied Optics {\bf 32}, 2976 (1993)

\bibitem{comment2} Our estimates based on $\sigma_{dc}$ suggest that
IR radiation probes the crystal to a depth of at least 5-10 $\mu$m.

\bibitem{comment1} Due to the unusual phonon structure in the spectrum,
we have not been able
to complete Kramers-Kronig analysis for CeFe$_4$P$_{12}$ samples
(we obtained physically unrealistic negative values for conductivity).
Instead, we fitted the reflectance
spectra and, using the same parameters, generated the conductivity
and dielectric constant spectra which are shown as thin lines in Fig.4 and 6.
The free-electron part was fitted with the Drude formula and
the phonons were modeled as Lorenzian oscillators.

\bibitem{tanner93} D.B.Tanner and T.Timusk in {\it Physical Properties of
High Temperature Superconductors III}, edited by D.M.Ginsberg
(World Scientific, 1993)

\bibitem{grandjean84} F.Grandjean, A.Gerard, D.J.Braun and W.Jeitschko,
J.Phys.Chem.Solids Vol. {\bf 45}, No.8/9, 877-86, (1984)

\bibitem{danebrock96} M.E.Danebrock, C.B.H.Evers and W.Jeitschko,
J.Phys.Chem.Solids, Vol.{\bf 57}, No.4, 381-7, (1996)

\bibitem{evers95} C.B.H.~Evers, W.~Jeitschko, L.~Boonk, D.J.~Braun, T.~Ebel
and U.D.~Scholz, Journal of Alloys and Compounds, {\bf 224}, 184-9, (1995)

\bibitem{singh96} D.J.~Singh, L.~Nordstrom, W.E.~Pickett and J.L.~Feldman,
15$^{th}$ International Conference on Thermoelcetics (1996)

\bibitem{singh97} D.J.~Singh and I.I.~Mazin, Phys.Rev.B, {\bf 56},
R1650-3, (1997)

\bibitem{nordstrom96} L.~Nordstr\"{o}m and D.J.~Singh, Phys.Rev.B
{\bf 53} 1103-8 (1996)

\bibitem{ogut96} S.~\"{O}\u{g}\"{u}t and K.M.~Rabe, Phys.Rev.B {\bf 54},
No.12, R8297-300, (1996)

\bibitem{damascelli97} A.~Damascelli, K.~Schulte, D.~van der Marel and
A.A.~Menovsky, Phys.Rev.B, Vol.{\bf 55}, No.8, R4863-6, (1997)

\bibitem{falicov81} {\it Valence Fluctuations in Solids}, edited by
L.M.Falicov, W.Hanke and M.B.Maple (North-Holland, Amsterdam, 1981)

\bibitem{lutz82} H.D.~Lutz and G.~Kliche, Phys.Stat.Sol (b), {\bf 112},
549 (1982)

\bibitem{fateley72} W.G.~Fateley, F.R.~Dollish, N.T.~McDevitt and
F.F.~Bentley {\it Infrared and Raman Selection
Rules for Molecular and Lattice Vibrations: The Correlation Method}
(Wiley-Interscience, New York, 1972)

\bibitem{bruesch82} P.Br\"{u}esch, {\it Phonons: Theory and Experimenst}
(Springer, New York, 1982), Vols.1 and~2.

\bibitem{chen97} B.~Chen, J.~Xu, C.~Uher, D.T.~Morelli, G.P.~Meisner,
J.~Fleurial, T.~Caillat and A.~Borchevski, Phys.Rev. {\bf 55},
No.3., 1476-80, (1997)

\bibitem{keppens98} V.~Keppens, D.~Mandrus, B.C.~Sales, B.C.~Chakoumakos,
P.~Dal, R.~Coldea, M.B.~Maple, D.A.~Gajewski, E.J.~Freeman and S.~Bennington,
Nature, vol.{\bf 395}, 876-8, (1998)

\bibitem{mandrus97} D.~Mandrus, B.C.~Sales, V.~Keppens, B.C.~Chakoumakos,
P.~Dai, L.A.~Boatner, R.K.~Williams, J.R.~Thompson, T.W.~Darling, A.~Migliori,
M.B.~Maple, D.A.~Gajewski and E.J.Freeman, Mat.Res.Soc.Symp.Proc.,
Vol.{\bf 478}, 199-209, (1997)

\bibitem{shirotani99} I.~Shirotani, T.~Uchiumi, C.~Sekine, M.~Hori and
S.~Kimura, Journal of Solid State Chemistry, {\bf 142}, 146-51, (1999)

\bibitem{dordevic99} S.~Dordevic, N.R.~Dilley, D.N.~Basov, E.D.~Bauer and
M.B.~Maple, to be published, (1999)

\bibitem{feldman99} J.L.~Feldman, D.J.~Singh and I.I.~Mazin, to be published,
(1999)

\end{references}
\end{document}